\documentclass[prb,preprint,showpacs]{revtex4}
\usepackage{t1enc,epsfig,subfigure,epsf,graphicx,amssymb,verbatim,float}

\begin{document}
\bibliographystyle{apsrev}

\title{A Structural Model for Octagonal Quasicrystals Derived from Octagonal
Symmetry Elements Arising in $\beta$-Mn Crystallization of a
Simple Monatomic Liquid}

\author{M{\aa}ns Elenius, Fredrik~H.~M.~Zetterling and Mikhail Dzugutov}
\affiliation{Department of Numerical Analysis and Computer Science,
             Royal Institute of Technology, SE--100 44 Stockholm, Sweden}
\author{Daniel~C. Fredrickson and Sven Lidin}
\affiliation{Department of Inorganic Chemistry, Arrhenius Laboratory,
             Stockholm University, SE--106 91 Stockholm, Sweden}

\date{\today}
\begin{abstract}

While performing molecular dynamics simulations of a simple monatomic liquid, we observed the crystallization of a material displaying octagonal symmetry in its simulated diffraction pattern.  Inspection of the atomic arrangements in the crystallization product reveals large grains of the $\beta$-Mn structure aligned along a common 4-fold axis, with 45$^\circ$ rotations between neighboring grains.  These 45$^\circ$ rotations can be traced to the intercession of a second crystalline structure fused epitaxially to the $\beta$-Mn domain surfaces, whose primitive cell has lattice parameters $a$ = $b$ = $c$ = $a_{\beta-Mn}$, $\alpha$ = $\beta$ = 90$^\circ$, and $\gamma$ = 45$^\circ$. This secondary phase adopts a structure which appears to have no known counterpart in the experimental literature, but can be simply derived from the Cr$_3$Si and Al$_3$Zr$_4$ structure types.  We used these observations as the basis for an atomistic structural model for octagonal quasicrystals, in which the $\beta$-Mn and the secondary
phase structure unit cells serve as square and rhombic tiles (in projection), respectively. Its diffraction pattern down the octagonal axis resembles those experimentally measured. The model is unique in being consistent with high-resolution electron microscopy images showing square and rhombic units with edge-lengths equal to that of the $\beta$-Mn unit cell. Energy minimization of this configuration, using the same pair potential as above, results in an alternative octagonal quasiperiodic structure with the same tiling but a different atomic decoration and diffraction pattern.

\end{abstract}


\maketitle

\section{Introduction}

Quasicrystals are materials whose diffraction patterns show what at first glance is a striking contradiction:  the sharp peaks normally arising from long-range periodicity, arranged in patterns of icosahedral, decagonal, dodecagonal or octagonal symmetries---all fundamentally incompatible with periodic ordering.  Since the discovery of these puzzling and wonderful objects,\cite{shechtman84prl} significant progress has been made in characterizing the atomic arrangements underlying their diffraction patterns.  For instances of both the icosahedral and decagonal families, detailed crystal structures have been solved from X-ray diffraction data,~\cite{tsai07natmat,steurer07jac} while the solution of a series approximant structures to the Ta$_{1.6}$Te quasicrystal, also from X-ray diffraction experiments, has brought insights into the atomic geometries in the dodecagonal family.~\cite{harbrecht98ac, harbrecht00mse, harbrecht02cej}

In contrast, only limited structural information is available for the octagonal family of quasicrystals.  Just a handful of examples have been experimentally realized, those in the V-Ni-Si, Cr-Ni-Si, Mn-Si and Mn-Si-Al systems.~\cite{Kuo87,kuo88apl, Steurer04}  In all of these cases, their preparation involves the rapid cooling of an alloy melt, a method not conducive to the formation of high quality crystals for structural analysis with X-rays.  Prolonged heating, the most likely path to the growth of such crystals, inevitably leads to decomposition of the quasicrystal, usually into a $\beta$-Mn-type phase,~\cite{Kuo87,kuo88apl,kuo89prb} but the formation of a AuCu$_3$-type phase has also been observed.~\cite{xu00prb}  Because of this metastability, the techniques for investigating these phases experimentally have been limited mainly to high resolution electron microscopy (HREM) and electron diffraction.

Such studies revealed that these structures can be viewed as octagonal tilings of square and rhombi units (with 45$^\circ$ angles at their acute corners), which is one of several possible tilings giving octagonal symmetry,~\cite{socolar89,ingalls93jncs} and can be embedded into higher-dimensional spaces using superspace methods.~\cite{kuo88aca,li96cpl} It was also observed that the edges of these units are metrically equal to that of a $\beta$-Mn-type structure.~\cite{Kuo87}  However, the positions of individual atoms have not been thus far resolvable.  This gap between the experimental investigations of octagonal quasicrystals and their detailed structures at the atomic level places obvious limits on our ability calculate theoretically their physical properties.  In attempts to calculate their electronic structure~\cite{zijlstra04} and vibrational properties,~\cite{liu92jpcm} researchers were forced to resort to an unrealistic model in which a single atom is placed at every vertex of an octagonal tiling of squares and rhombi.

Several attempts have been made to fill this gap.  Two different structure models, those of Kuo et al.~\cite{kuo90jncs,kuo91prl,li96cpl} and Hovm{\"o}ller et al.~\cite{hovmoeller91pml} have been proposed making use of the apparently close structural relationship between the octagonal quasicrystals and the $\beta$-Mn structure.  Both models extract square and rhombus units from the $\beta$-Mn structure, and then use these as tiles in an octagonal tiling, the difference being in the definition of the square and rhombic units.  While these models exhibit diffraction patterns with similarities to the experimentally measured ones, they are incompatible with the  tile-edge lengths measured from HREM images.  A third model proposed by Hovm{\"o}ller et al. derived from 3D reconstructions of HREM images also is inconsistent with this observation.~\cite{hovmoeller95pml}

In this paper, we use the results of molecular dynamics simulations in an attempt to bring understanding of the structures of octagonal quasicrystals.  This approach has been successfully employed by ourselves and others to observe the formation of dodecagonal and decagonal quasicrystals from monatomic liquids.~\cite{Dzugutov93,Quandt99,engel07prl,keys07prl}

In our continued work with such simulations, we came across a crystallization product which showed an octagonal symmetry axis in its calculated diffraction pattern.  In the course of this paper, we will examine the origins of this octagonal diffraction pattern in the atomic arrangements of the sample.  As we will see below, this octagonal symmetry arises from a 45$^\circ$ difference in orientation between large grains of the $\beta$-Mn structure type, rather than the occurrence of a true octagonal phase.  However, the geometrical reasons for these 45$^\circ$ relative rotations form the roots of a new structural model for octagonal quasicrystals. This model affords not only a diffraction pattern with correspondence to the experimental ones down the octagonal axis, but also is unique in having the $\beta$-Mn unit cell repeat period as the length of its square and rhombus units, in accord with HREM images.  Central to this construction are 8$_3$ screw helices implemented in Hovm{\"o}ller et al.'s 1991 model~\cite{hovmoeller91pml}---this time used in a different decoration of the square and rhombic tiles.

\section{$\beta$-MN Formation in Molecular Dynamics Simulations}

Our investigations into the structures of octagonal quasicrystals began with some simulations of a monatomic liquid, using a pair potential designed to encourage icosahedral ordering.  We started this simulation by first equilibrating a system of atoms in a high temperature liquid state. We then began a stepwise cooling at constant density ($\rho$= 0.84 atoms/unit volume, in reduced units, see Appendix) with intermediate equilibration. At a point in this cooling, the system showed a marked drop in potential energy, pressure and diffusivity, indicating crystallization (for further details of the simulation, see Appendix).

\begin{figure}
      \includegraphics[width=9.cm,trim=2cm 0 2cm 1.5cm,clip]{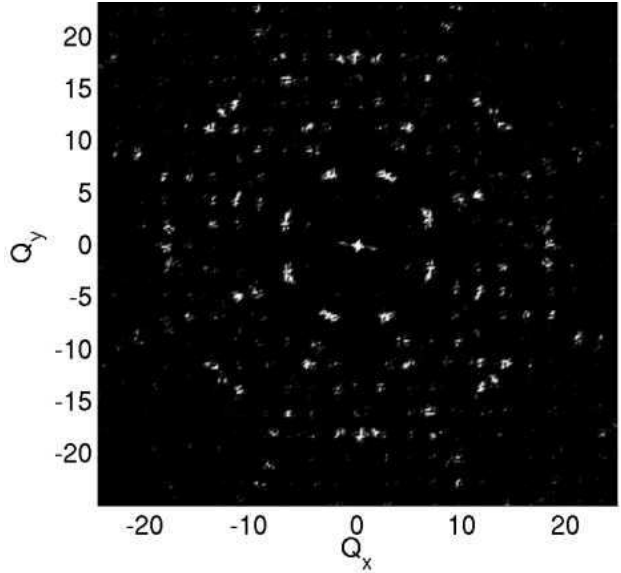}
      \caption{Calculated diffraction intensities for our simulation crystallization product in the plane perpendicular to the octagonal axis.}
  \label{diff}
\end{figure}

To identify the crystallization product, we first analyzed the radial distribution function $g(r)$ and the structure factor $S(Q)$ of the sample. These confirmed that the structure had relatively long-range order, indicative of a crystalline structure. We then turned to a more detailed analysis of the diffraction pattern. The diffraction intensities for a configuration of N point particles are calculated as:
\begin{equation}
S({\bf Q}) = N^{-1} \left |\sum_{i=1}^N e^{i{\bf Q r}_i} \right |^2
\end{equation}
where vectors ${\bf r}_i$ denote the particle positions. First we determined the orientations of the crystallographic axes of the phase in reciprocal space. To do this, we plotted the values of S({\bf Q}) on the sphere in reciprocal space corresponding to the maximum of $S(Q)$. What we found was surprising: a symmetry axis of eight-fold symmetry, with eight perpendicular two-fold axes.  Having found this octagonal symmetry axis, we then calculated diffraction intensities in the plane perpendicular to it (Figure \ref{diff}). In this reciprocal space plane, the eightfold symmetry is apparent. Hence we arrive at the, now classical, contradiction seen in quasicrystals of sharp diffraction spots indicative of long-range order and rotational symmetry incompatible with periodic ordering.

\begin{figure}
  \includegraphics[width=14.cm]{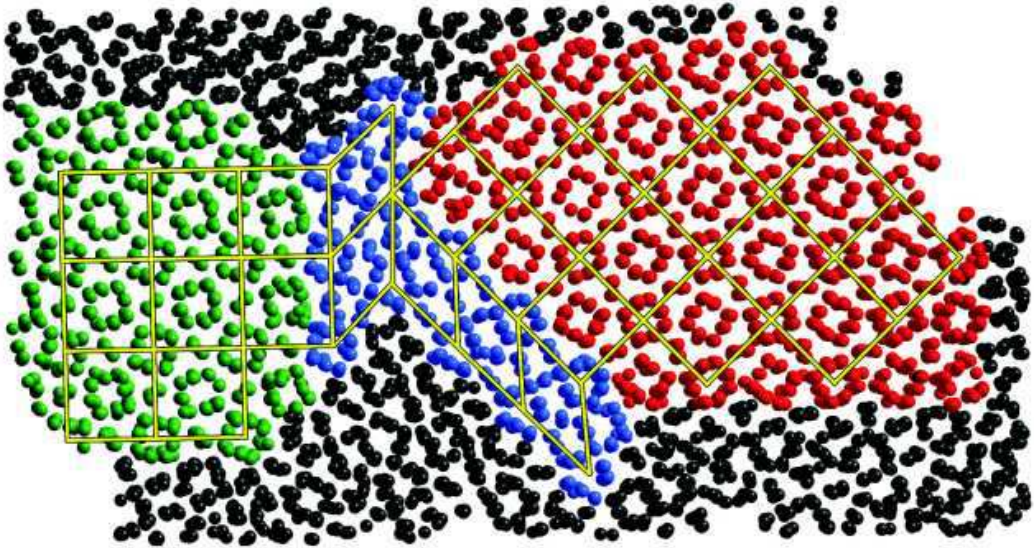}
  \caption{(Color)  Two regions of the $\beta$-Mn structure (green and red) and their
  surroundings in the crystallized sample resulting from our molecular dynamics simulation (see text). Perpendicular to the plane of the page, the crystallographic order of these grains propagates infinitely (the sample periodic along this direction, with a repeat period corresponding to eight $\beta$-Mn cells).  Small fragments of a second crystalline phase can be perceived between these grains (blue).  Yellow lines trace out unit cells of these two structures.  As the secondary phase unit cell appears as a nearly 45$^\circ$ rhombus in this projection, we'll refer to this as the rhombic phase structure.}
  \label{rspaceall}
\end{figure}

This encouraged us to look at the real space atomic configuration.  Instead of the expected octagonal atomic arrangements, we found large domains of a crystalline structure.  Figure \ref{rspaceall} shows approximately 7000 atoms in the simulation box, with two grains of this structure highlighted in green and red, respectively.  Unit cell edges for these crystallites are drawn in with yellow squares.  Further inspection of these grains revealed them to be of the $\beta$-Mn structure type.  A look at these cell edges reveals that the crystallographic axes of the two grains have different orientations.  In the domain marked in green, these are aligned with the horizontal and vertical directions of the page.  In the red one, however, they are oriented diagonally.  The green and red domains are, in fact, rotated by 45$^\circ$ relative to each other.  This 45$^\circ$ rotation accounts for the 8-fold character of the diffraction pattern calculated for this geometry.

The occurrence of these beta-Mn grains together with the interstitial spaces between them correlates with density heterogeneities, the crystalline grains coinciding with the denser regions.  This result can be rationalized in terms of energetics: model calculations show that the $\beta$-Mn structure is an energy minimum for this potential, with the ideal density being $\rho$=0.96, compared to $\rho$=0.84 in the simulation.  Indeed, the difference in unit cell size between the $\beta$-Mn cells in the sample and that of the ideal crystal at $\rho=0.96$ is less than $1\%$.

The small $\beta$-Mn crystallites are not the only geometrical regularity that can be recognized in Figure~\ref{rspaceall}.  At the edges of the green and red domains, near the center of the figure, two sets of cross-linked rows of atoms are highlighted in blue.  Similar rows recur frequently as terminating features at the $\beta$-Mn domain surfaces.  As shown with yellow lines overlaid on these atoms in blue, a unit cell can be identified for this pattern.  In other words, these surface features correspond to small domains of a second crystalline phase in the simulation result. The unit cell for this phase appears as a rhombus in this projection, and for this reason we will refer to it as the rhombic phase.  As we will see in the next section, a closer inspection of this structure provides insights into the prominence 45$^\circ$-twinning in this sample, and the relationship between the $\beta$-Mn structure and octagonal quasicrystals.

\section{A second phase: rhombi with the squares}

\begin{figure}
  \includegraphics[width=9.cm]{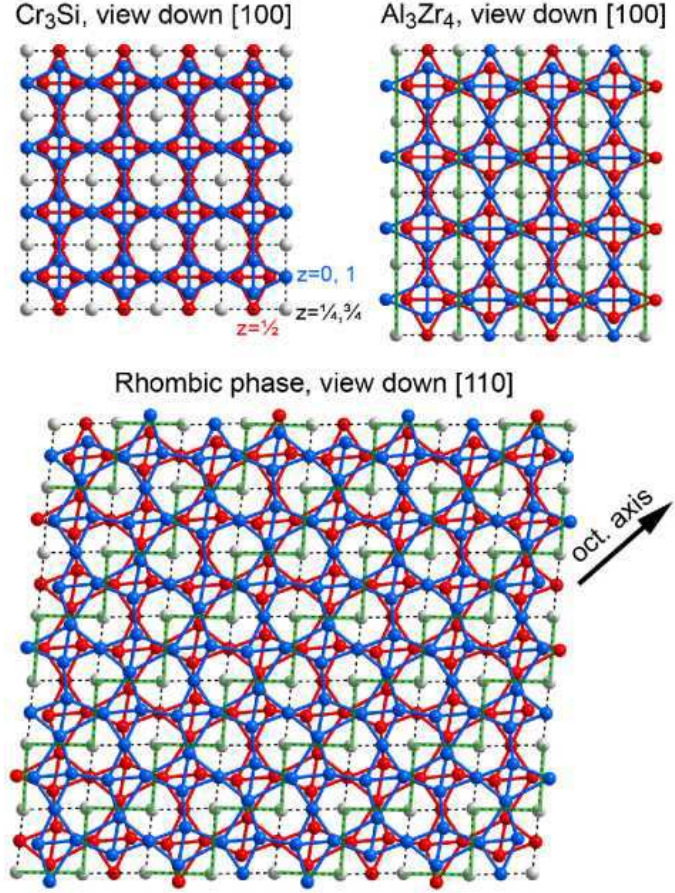}

  \caption{(Color) Structural relationships between the second phase detected alongside $\beta$-Mn in molecular dynamics simulations (the rhombic phase), and two well-known intermetallic structure types: those of Cr$_3$Si and Al$_3$Zr$_4$.  In these projections, atoms in red and blue are at heights 0 and 1/2, respectively.  Those in light gray occur at both heights 1/4 and 3/4.  Green bars: shear planes across which regions of local Cr$_3$Si structure type geometry are interrupted by shifts in height of 1/2, with the deletion of some atoms which are brought into near coincidence with others following these shifts. A black arrow shows the orientation of the octagonal axis occurring in the simulation results.}
  \label{rphasecomp}
\end{figure}

\begin{figure}
  \includegraphics[width=14.cm]{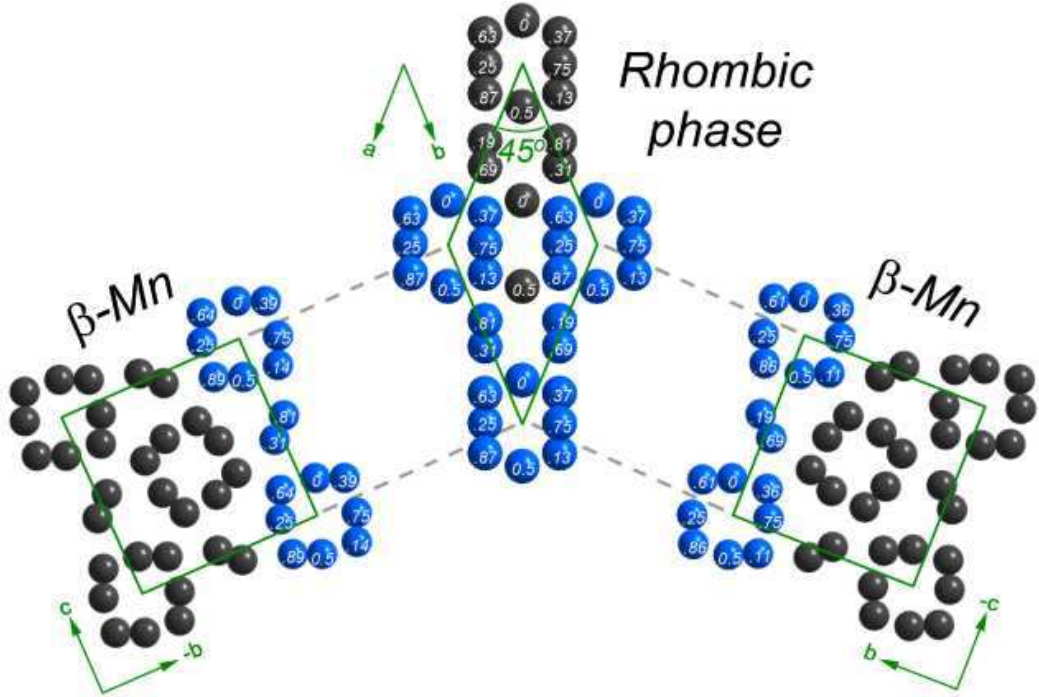}
  \caption{(Color Online)  Epitaxial matching between the \{100\}
  planes of the $\beta$-Mn structure and the (100) and (010) planes of
  the rhombic phase structure.  $\beta$-Mn grains interfacing at these
  two different planes of the rhombic phase differ in their
  orientation by 45$^\circ$.  Atoms to be merged at the interface are
  drawn in blue in the online version.}
  \label{interface45}
\end{figure}

Upon identifying the rhombic phase, we read off approximate atomic coordinates from the positions in the simulation result and generated a fully periodic model.  Prompted by the close match in cell parameters to those of the $\beta$-Mn structure, and the near 45$^\circ$ angle of the rhombus, we idealized the unit cell to $a$ = $b$ = $c$, $\alpha$ = $\beta$ = 90$^\circ$, $\gamma$ = 45$^\circ$. We then performed a steepest descent energy minimization to obtain more accurate atomic positions.  The resulting rhombic phase structure is shown in Figures~\ref{rphasecomp} and \ref{interface45}.

As far as we can tell, this crystal structure has no counterpart in the experimental literature. However, it shows structural similarities to the Frank-Kasper family of structures, and can be viewed as a derivative of the Cr$_3$Si and Al$_3$Zr$_4$ structure types.  This is illustrated in Figure \ref{rphasecomp}, in which we show that the structures of both Al$_3$Zr$_4$ and the rhombic phase in our simulations can be derived by introducing variations into the Cr$_3$Si structure type.  The Cr$_3$Si-type can be viewed (along with many other ways) as a stacking of nets:  nets built from hexagons and triangles (red and blue) alternate with more open square ones (gray spheres, dotted lines).  The Al$_3$Zr$_4$ structure, when viewed down the [100] direction, shows similar features.~\cite{anderssonbook}  In fact, a simple way to generate the Al$_3$Zr$_4$ structure is to begin with the Cr$_3$Si-type and introduce shear planes in rows parallel to one set of lines in the Cr$_3$Si-type's square nets, as shown with green lines in the top right corner of Figure \ref{rphasecomp}.  Across each of these planes we introduce a shift of half of a unit cell out of the page (accompanied by the deletion of some atoms that come into near coincidence with atoms on the other side of the shear plane following the shift).  If we introduce one such shear plane for every row of squares in the projection of Figure \ref{rphasecomp}, we arrive at the Al$_3$Zr$_4$ structure type.

The rhombic phase in our simulations can be derived from the Cr$_3$Si structure in a similar manner.  At the bottom of Figure \ref{rphasecomp}, we show a view of the rhombic phase structure perpendicular to that of Figure \ref{rspaceall}.  In this view, similarities to the Cr$_3$Si structure can be seen in the square nets (with some distortion) and the filling of these squares with atoms in red and blue.  The details of the structure can be derived, just as for the Al$_3$Zr$_4$ structure type, by inserting shear elements into Cr$_3$Si-type parent structure.  While in the Al$_3$Zr$_4$ structures these shear elements occur as regularly spaced planes parallel to one set of lines in the square nets, in the rhombic phase the shears occur diagonally across the square net, moving through the square nets in a zigzag pattern.  The rhombic phase structure can be seen then as a variation on the Al$_3$Zr$_4$ structure.

A look at this structure from a different viewpoint, down the same direction as in Figure \ref{rspaceall}, helps us to understand why this phase so frequently appears at the surfaces of $\beta$-Mn crystallites in the simulation.  In Figure \ref{interface45} we illustrate this for a single unit cell of the rhombic phase, with the heights of the atoms given with white numbers (in fractions of the unit cell repeat vector coming out of the page).  In this view, the rhombic phase unit cell appears as a rhombus with circular 8-atom wreaths of atoms at each corner (and the very center of the cell). A closer inspection of the heights of the atoms in one of these wreaths reveals that they trace out helices coming out of the plane of the page, with a local 8$_3$ symmetry axis passing through the center of the ring.

These 8$_3$ helices have close counterparts in the $\beta$-Mn structure.  In the left and right of Figure \ref{interface45} we draw unit cells of the $\beta$-Mn with their unit cell edges aligned with those of the rhombic phase cell.  The corners of the $\beta$-Mn cells show similar helices as in the rhombic phase.  In fact, if we compare the heights of the atoms at the cell edges of the two structures types, we find a one-to-one correspondence of atoms at nearly identical sites.  A nearly perfect epitaxial mapping exists between the sides of the $\beta$-Mn unit cell, and those of the rhombic phase.  It is not surprising, then, that grains of $\beta$-Mn should be terminated by layers of the rhombic phase.

This epitaxial mapping can be used to understand the prevalence of $\beta$-Mn grains oriented relative to each other by a 45$^\circ$ rotation.  Note from Figure \ref{interface45} that the rhombic phase can interface with $\beta$-Mn crystallites at both its (100) and (010) sets of planes.  As the $\gamma$ angle of this unit cell is 45$^\circ$, these two sets of planes are also inclined relative to each other by 45$^\circ$.  A $\beta$-Mn crystallite interfacing with a (100) plane will thus be misaligned by 45$^o$ relative to another crystalline interfacing with a (010) plane.  We posit that the 45$^\circ$-twinning in our simulations results from the presence of this rhombic phase in the interstices between $\beta$-Mn grains.

Several factors in the energetics of the $\beta$-Mn and rhombic phase structures elucidate the observations we've made in our simulations.  The $\beta$-Mn is energetically preferable with the current potential, consistent with the fact that we see more atoms in the $\beta$-Mn domains than in the rhombic phase one in our simulation results.  More importantly, the two structures have energy minima at nearly equal densities ($0.96$ for $\beta$-Mn, $0.95$ for the rhombic phase), and these densities give almost identical unit cell lengths, making intergrowth of the two structures particularly facile.

With the ease with which unit cells of $\beta$-Mn and the rhombic phase fit together, it's tempting to imagine other ways of arranging them in space to generate new structures.  After all, octagonal quasicrystals are usually viewed as tilings of squares and rhombi just like these.  It's indeed tempting, and, as we shall see in the following sections, this is a temptation we can't resist.

\section{Tiling with the structures of $\beta$-Mn and the rhombic phase}

In the previous two sections, we identified two crystalline structure types in the results of our molecular dynamics simulations of a simple monatomic liquid, those of $\beta$-Mn and a previously unobserved structure, which we'll call the rhombic phase.  The unit cell dimensions and atomic arrangements in these two structures are propitious to the formation of epitaxial interfaces between them.  In the simulations, this is reflected in the appearance of rhombic phase layers terminating grains of the $\beta$-Mn structure.  Many more possibilities become evident when we imagine the unit cells of these structures as square and rhombic tiles with which we can cover the plane (taking the periodic stacking of the cells in the third dimension as a given), the square and rhombic tiles being derived from the $\beta$-Mn and rhombic phase structures, respectively.  In the following paragraphs, we will show that the unit cells of these two structures are compatible with any tiling of squares and rhombi.

\begin{figure}
  \includegraphics[width=9.cm]{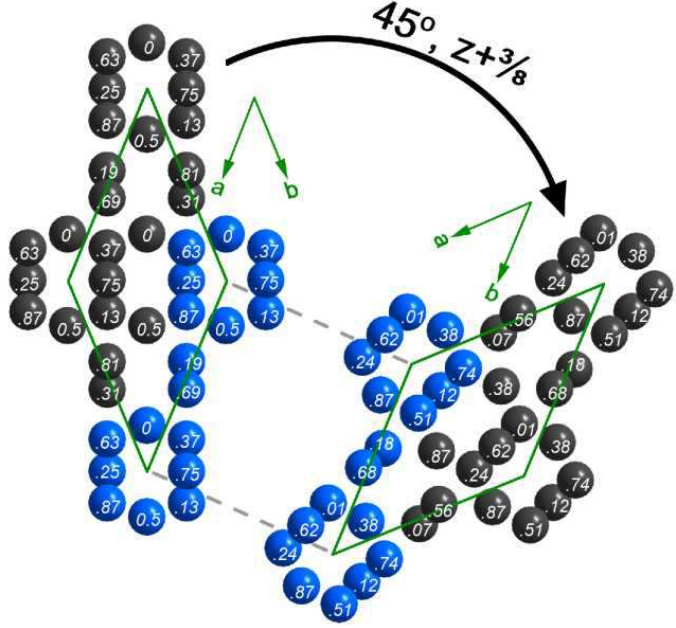}
  \caption{(Color Online) Epitaxial interface between one rhombic
  phase unit cell, and a second cell generated from the first using an
  8$_3$ symmetry operation. Atoms to be merged at the interface are
  drawn in blue in the online version.}
  \label{interface45partII}
\end{figure}

Several interface types are geometrically feasible using these two tiles. We've already seen that the squares join naturally at the corners and edges with the rhombi (corners and edges referring to the 2D projection in Figure~\ref{interface45} and those to come; in 3D, they are cell edges and faces, respectively).  Interfaces between two square tiles and between two rhombic tiles are seen, of course, in the crystal structures from which they are derived.  In Figure~\ref{interface45partII} we show a fourth interface type, between two rhombic tiles differing in orientation by 45$^\circ$.  The local pseudo 8$_3$ symmetry axes passing through the corners of the rhombic phase unit cell makes it possible to fuse two such tiles.  The two tiles simply need to be related by an 8$_3$ operation, i.e. the 45$^\circ$ difference in orientation is accompanied by a 3/8 shift in the heights of the atoms between tiles.  As can be seen in the figure, applying such a 8$_3$ operation to one rhombic tile creates a second tile which is well-suited to edge-sharing with the original one.  While there is some degree of mismatch in the coordinates within the plane, due to the approximate nature of the 8$_3$ axis, there is a one-to-one correspondence between the atoms at the edge to be shared between the two tiles.  Also, all corresponding atoms at the edge match in height within one percent of the unit cell repeat distance out of the plane.  In summary, the pseudo 8$_3$ symmetry of the rhombic tiles' corners means that two rhombic tiles related by an 8$_3$ symmetry element at a corner can be edge-fused together.

Similar geometrical arguments can be used to derive a simple matching rule for building larger structures with square and rhombic tiles.  Both tiles have  pseudo 8$_3$-symmetry axes at their corners.  This means that when two tiles come together, they can share corners or edges, as long as their difference in orientation is accompanied by an appropriate relative translational shift of their atoms out of the plane.  Specifically, clockwise rotations of a tile by 45$^\circ$ must be accompanied by translations in height of 3/8.  These translational shifts align the phases of the helices at their corners so that they meet in register.  Adherence to this rule can be seen in the interface figures presented so far, Figures~\ref{interface45} and~\ref{interface45partII}.  In both figures, each 8$_3$ helix at a tile corner occurs with the  same orientation and phase.

\begin{figure}
  \includegraphics[width=9.cm]{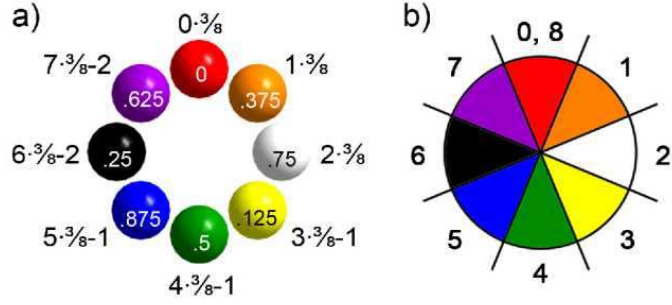}
  \caption{(Color)  A simplified notation for the 8$_3$ helices occurring in the $\beta$-Mn and rhombic phase structures.  (a) an idealized 8$_3$ helix projected down its helical axis.  Heights for the atoms are given in both decimal notation and as multiples of 3/8. (b) A color wheel representation of this helix.  The wheel is divided into eight wedges, each color-coded according to the height of the atom occurring in the corresponding octant in the projection of (a). }
  \label{colorwheel}
\end{figure}

The key role played by the 8$_3$ axes in this matching rule suggests a simple notation for describing structures based on these tiles.  In  Figure~\ref{colorwheel}a, we show an idealized 8$_3$ helix, viewed down its long axis, with the heights of the atoms drawn in both decimal and fraction notations.  The height of each atom can be expressed as a different multiple of the fraction 3/8.  We can simplify this picture by representing the helix as a circle divided into 8 equal slices (Figure~\ref{colorwheel}b), with each slice color-coded according to the height of the atom occupying that octant in the view of Figure~\ref{colorwheel}a.

\begin{figure}[htbp]
  \includegraphics[width=14.cm]{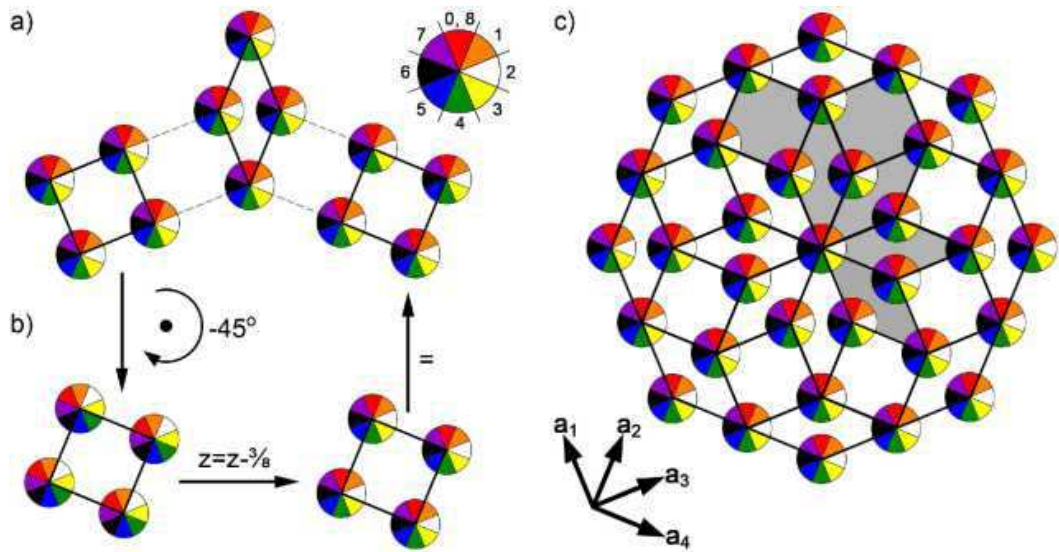}
  \caption{(Color)  Illustration of tiling a space with cells of the $\beta$-Mn and rhombic phase structures, appearing respectively as square and rhombic tiles in projection.  (a) Simplified representation of the interfaces in Figure \ref{interface45}, in which the 8$_3$ helices at the tile corners are drawn with the color-wheel notation of Figure \ref{colorwheel}.  The remaining atoms in the cells are not shown in this representation for the sake of  clarity. (b) A graphical demonstration that the two square tiles in (a) can be generated from each other via 8$_3$ symmetry operations. (c) An octagonal cluster built from $\beta$-Mn and rhombic phase tiles.  Six distinct tiles occur in this pattern; one instance of each is shaded in gray. This octagonal cluster can be generated using the four basis vectors given in the lower left corner.  See text for a proof that the six distinct tiles indicated here can be used to decorate any tiling of squares and rhombi.}
  \label{tilingwheels}
\end{figure}

An example of how this notation can be used in structural descriptions is given in Figure~\ref{tilingwheels}a, in which we take another look at the $\beta$-Mn/rhombic phase/$\beta$-Mn interface of Figure~\ref{interface45}.  Here, rather than seeing simple geometrical figures decorated by complex arrangements of atoms at various heights, we now see a rhombus meeting two squares, with colorful wheels at all of the corners.  At both interfaces the circles at the corners match in their color patterns, meaning that they can merge in a facile manner.

The color-wheel notation also simplifies the investigation of relationships between tiles in different orientations.  In Figure~\ref{tilingwheels}b, we show a simple graphical demonstration that the two square tiles are related by a 8$_3^{-1}$ operation.  This symmetry relationship means that while the color patterns differ within the square interiors of these two tiles, they represent the same crystal structure. They differ only in their orientations and by a relative shift in their atomic heights.  We see, then, that this particular arrangement of tiles is compatible with a simple intergrowth of the $\beta$-Mn and rhombic phase structure types.  No new geometrical arrangements have been forced within the tiles.

More complex patterns of squares and rhombi can be created, and their feasibility checked, using the same notation. In Figure~\ref{tilingwheels}c, we've generated an octagonal cluster, starting at the center with an eight-fold star built from eight rhombi sharing a central point. The niches of this star were then filled-in with squares, with the nascent joints serving as sockets for additional rhombi.  The atomic geometries required by this octagonal figure can be probed by placing color-wheels, again representing 8$_3$-helices, at each corner with the same phase, and inspecting the resulting tile colorings.  As can be seen by inspection of the tiles, the entire pattern is built up from only six distinct tile types. One example of each is shaded in with gray.  Two are square tiles, which are identical to those found in Figure~\ref{tilingwheels}a-b.  As we saw above, they both represent a single unit cell of the $\beta$-Mn structure.  The remaining four are rhombic tiles.  The first of these is oriented vertically as in the central rhombus of Figure~\ref{tilingwheels}a, and the remaining three are generated from this by the application of one, two, and three 8$_3$ operations.  As each of these rhombic tiles are related to the others by  8$_3$ operations, we can see that their decorations are single unit cells of the same structure type, the rhombic phase type.  No new geometrical arrangements in the atomic decorations of the tiles have been enforced by this tiling pattern.  Furthermore, as these 8$_3$ operations leave the color patterns on the color-wheels invariant, these rhombic tiles join naturally at their corners and edges.

From this octagonal figure, we can generalize these observations to show that any tiling pattern of squares and rhombi can be decorated by single unit cells of the $\beta$-Mn and rhombic phase structure types.  This can be done in three steps:  (1) We note that while the pattern is fairly intricate, inspection of the pattern shows that it is based on only four basis vectors: ${\rm{\bf a_1}}$, ${\rm{\bf a_2}}$, ${\rm{\bf a_3}}$, and ${\rm{\bf a_4}}$ as drawn in the lower left corner of Figure~\ref{tilingwheels}c.  Starting with the central color-wheel, the full octagonal pattern can be generated by taking linear combinations of these basis vectors. (2) In our second step, we observe that there are a limited number of squares and rhombi that can be generated by making rings with these translation vectors.  They are in fact those six distinct tiles that we have highlighted in Figure~\ref{tilingwheels}c.  Any tiling of squares and rhombi traced out by linear combinations of these four basis vectors will contain only these six tiles.  (3) In our third and final step, we recognize that these basis vectors are sufficient to create any tiling of squares and rhombi that we may desire.

To summarize these arguments, we have found that the pseudo 8$_3$ symmetry elements at the corners of the unit cells of both the $\beta$-Mn and rhombic phase structures allow us to use these as square and rhombic tiles, respectively, with which we can tile the plane in any arrangement.  This would of course include quasiperiodic patterns such as those indicated by the electron microscopy studies of V-Ni-Si, Cr-Ni-Si, and Mn-Si-Al alloys by Kuo and coworkers.  Our $\beta$-Mn/rhombic phase intergrowth tilings provides the possibility of proposing atomistic models for such alloys to supplement the limited resolution of such microscopy investigations.  We will explore this possibility in the next and last major section of this paper.

\section{Quasicrystal modelling}

\begin{figure}[htbp]
  \includegraphics[width=14.cm]{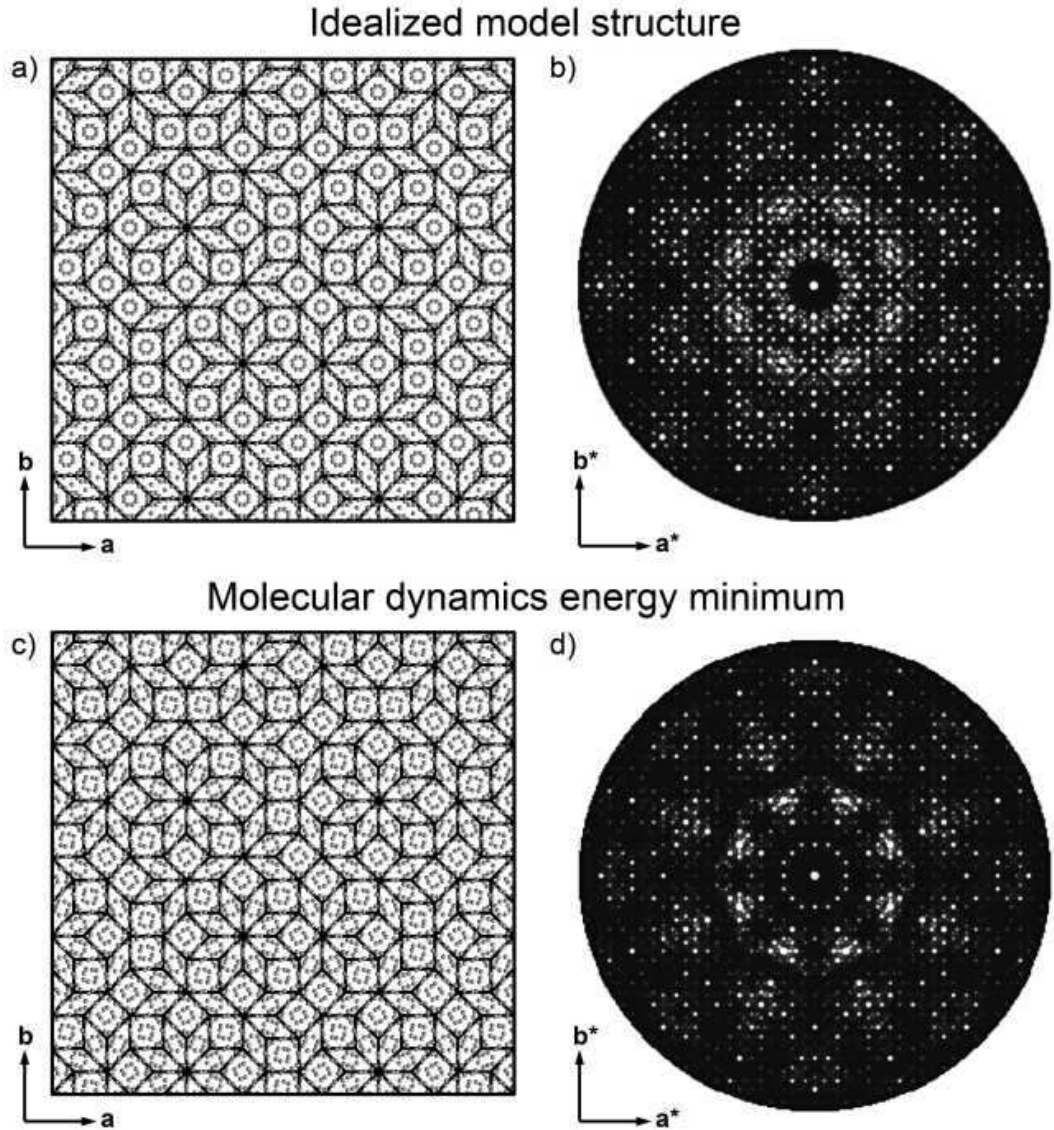}
  \caption{Octagonal quasicrystal approximants built from cells of the $\beta$-Mn and rhombic phase structures using the considerations of Section IV (see text), and their simulated diffraction patterns. (a) Idealized structure.  (b) A simulated electron diffraction precession photograph of the $hk0$ layer calculated for this structure. (c) The structure resulting from a steepest descent energy minimization of the structure in (a).  (d) The simulated diffraction pattern for this minimized structure. }
  \label{qcstructure}
\end{figure}

Octagonal tilings using rhombi and squares have already been extensively described.~\cite{socolar89} Such tilings rely on specific rules that enforce an aperiodic octagonal tiling. They also include inflation rules making it possible to go from a given tiling to one with more tiles per unit area, while preserving the aperiodicity and octagonal nature. Of particular importance for our work, which uses simulation programs requiring periodic structures,  is the observation that applying these inflations to periodic approximant structures leads to better and better approximants to the full aperiodic tiling.~\cite{Duneau89}  Zijlstra has constructed one periodic approximant which is well-suited to this application;~\cite{zijlstra04} it has the advantage of containing only one vertex where the tiling rules leading to quasiperiodicity are broken.  In the following we will use a one step inflation of this tiling as the basis for the construction of a structural model of octagonal quasicrystals.  This inflation gives a tiling of 239 tiles per unit cell, as seen in Figure~\ref{qcstructure}a, in which the tiles are delineated with black edges.

Having constructed an approximate octagonal tiling of squares and rhombi, we now decorate the tiles following the considerations of the previous section.  We place unit cells of the rhombic phase and $\beta$-Mn structure on the rhombic and square tiles, respectively, with the appropriate translational shifts to preserve the local 8$_3$ symmetry at the tile corners.

This decoration leaves room for significant freedom in choosing the exact atomic positions within the tiles, particularly if we remove the requirement that these cells obey the symmetries of their native lattices.  The only requirement is that the $8_3$-screws at the vertices and dumbbell atoms at each tile edge are preserved. We followed two main routes when selecting the exact atomic coordinates. In one, we tried to optimize the fit to the experimental diffraction pattern.~\cite{kuo91prl}  In doing this, we assumed perfect $8_3$-screws at each vertex and dumbbells centered at the edges; we then varied the internal distances in and rotations of these elements, while making sure reasonable interatomic distances were preserved.  We will call the result of this the idealized decoration. In our second approach, we aimed at finding the structure with minimal energy with our potential.  To do this, we performed a steepest descent energy minimization on the atomic configuration resulting from the first approach.

The result of these two approaches can be found in Figure~\ref{qcstructure}, along with their simulated electron diffraction patterns (putting a Mn atom at each atomic position, and adjusting the length scale accordingly).~\footnote{Electron diffraction simulations were made using the program eMap: AnaliteX Crystallographic Computing Software, Version 1.0, produced by AnaliTEX, Stockholm, Sweden.  Diffraction pattern images were exported from this program in gray-scale mode, with dark spots on a light background.  A separate graphics program was used to invert this scheme to aid comparison with experimentally measured patterns.} In both diffraction patterns we have inferred twinning with a mirror along the vertical axis in the electron diffraction pattern to reconcile the prominence of chiral $8_3$-screws present in our model with the clear mirror symmetry apparent in the experimental diffraction patterns, and the determination of the point symmetry of a Mn-Si-Al octagonal quasicrystal as 8/$m$ or 8/$mmm$ through convergent beam electron diffraction measurements.~\cite{kuo88apl}

Note that while the atomic arrangements resulting from these two approaches show rather small differences (Figures~\ref{qcstructure}a and c), these differences cause major changes in the electron diffraction patterns (Figures~\ref{qcstructure}b and d).  The most striking difference is that the diffraction intensity in is rather uniformly distributed about octagonal axis before the energy minimization (Figure~\ref{qcstructure}b), but becomes  spoked in appearance afterwards (Figure~\ref{qcstructure}d).

This difference between the diffraction patterns can be understood by a more detailed comparison of the atomic arrangements before and after the energy minimization.  Over the course of the energy minimization, the circular form of the 8$_3$ helices in the idealized structure of Figure~\ref{qcstructure}a has morphed into more square like arrangements (in projection, Figure~\ref{qcstructure}c).  This squaring of the 8$_3$ helices is reminiscent of the corresponding helices in the $\beta$-Mn structure.  In fact, during the minimization each tile has returned to a form much closer to the original unit cells of $\beta$-Manganese and the rhombic phase. Clearly our potential favors the local environments of the $\beta$-Mn structure to those of the idealized decoration. Even so, the octagonal nature of the minimized structure is still apparent in both real and reciprocal space images; hence {\it our simple spherical potential gives a minimum for this complicated octagonal structure.}

We are now in a position to understand the spoked appearance of the diffraction pattern of the energy minimized structure (Figure~\ref{qcstructure}c).  The spokes of intensity in the pattern lie parallel to the square tile edges in the real space structure.  In the true $\beta$-Mn structure these edges would lie parallel to 4$_1$-screw axes, which are known to create systematic absences in diffraction patterns.  For instance, a 4$_1$ axis along $a$ leads to the rule that for reflections with $k$ = 0 and $l$ = 0, only reflections with $h$ = 4$n$ will give nonzero intensity.  In the idealized structure, our idealization procedure has destroyed the 4$_1$ symmetry elements perpendicular to the octagonal axis, allowing a more uniform distribution of intensity about this axis.  With the return of $\beta$-Mn type character in the relaxed structure, these 4$_1$ axes have returned, at least on a local level; thus the observed spoked appearance of Figure~\ref{qcstructure}d is not surprising.

\begin{figure}[htbp]
  \includegraphics[width=14.cm]{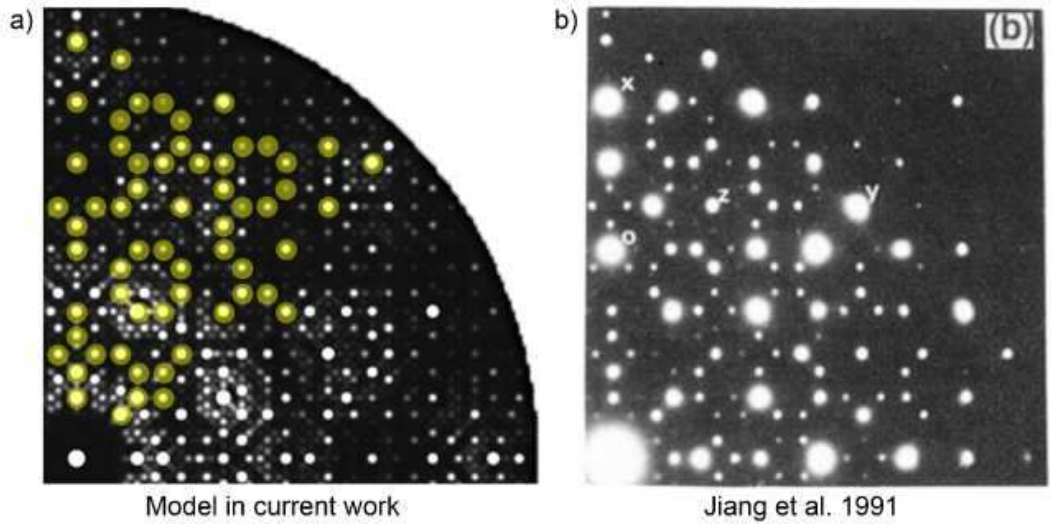}
  \caption{(Color Online) Comparison between the simulated electron
  diffraction pattern of (a) our idealized structure and (b) the
  pattern experimentally measured by Jiang et al. on a Mn-Si-Al
  octagonal quasicrystal. As a guide to the eye, some reflections in
  (a) corresponding to reflections in (b) are highlighted with yellow
  (light grey) circles. (b) is reproduced from the paper of Jiang et
  al. with the kind permission of the American Physical Society.}
  \label{diffdetails}
\end{figure}

As we described above, the idealized decoration was designed to give the best obtainable (by us) match to the experimentally measured electron diffraction patterns.  A detailed comparison between the simulated pattern for this structure to one of the highest quality electron diffraction images measured to date~\cite{kuo91prl} can be found in Figure~\ref{diffdetails}. Here, we use yellow spots to highlight key correspondences between the two patterns. While there are some differences between the two patterns,  the correspondence is still clear.
The chief differences are that (1) two peaks are missing, one rather strong outer peak and a less intense peak along the vertical axis and (2) some weak or moderate peaks present in the simulated pattern are not visible in the experimental pattern.  The extra peaks in the simulated pattern can easily be attributed to the fact that this pattern is calculated for a perfectly ordered structure; such perfection is not expected to occur in the experimental sample, particularly with the coarseness of the preparation method used.  The missing peaks are harder to explain, but could be attributed to dynamical scattering effects.

In fact, the fit between the simulation result and the experimental diffraction pattern is remarkable considering the simplicity of our model.  Further refinements could include attempts to go beyond an all-Mn occupation of the atomic sites, and allowing variations in the atomic decorations of the tiles based on their local environment.

\section{Conclusions}

In this paper, we began with the observation of the crystallization of $\beta$-Mn grains in a molecular dynamics simulation of a simple monatomic liquid.  A curious 45$^\circ$ twinning occurred between these grains, imparting an octagonal symmetry to the sample's calculated diffraction pattern.  This twin law could be traced to the presence of a secondary crystalline phase, whose unit cell appears as a 45$^\circ$ rhombus in projection, the sides of which interface cleanly with the faces of $\beta$-Mn unit cells.  This provided an explanation for the observed twin law, and also served as a basis for the construction of hypothetical structures using these unit cells as tiles.  We extended this to build a detailed structural model of an octagonal quasicrystal.  This model not only shows a reasonable match to the experimental electron diffraction patterns, but is also the only structural model thus far proffered that is consistent with the observation that the tile edges have the same length as the unit cell edge of the $\beta$-Mn structure.~\cite{Kuo87,Steurer04}

We found that energy minimization of this model structure results in significant changes to the tile decoration and diffraction pattern, but the overall tiling remains unchanged.  Energy minimization, then, yields an alternative octagonal structure model.  This finding---that a single-component system of particles interacting via a spherically symmetric potential exhibits a mechanically stable energy minimum configuration with an octagonal diffraction pattern---is of significant conceptual interest. It demonstrates that the structural complexity of octagonal quasiperiodic order experimentally observed in multicomponent intermetallic phases, with pronounced directional bonding, may be largely reduced to a simple monatomic archetypal quasicrystal.

We envision several avenues by which further insights into octagonal quasicrystals may be followed using molecular dynamics simulations.  The first would be the pursuit of a modified version of our pair potential that is prone to the crystallization of an octagonal phase.  One could also imagine studying possible decomposition routes for octagonal phases, by annealing samples of our ideal quascrystal structure model using the present potential.  Insights into the ways octagonal tilings are stabilized could be approached through the analysis of the energetic relevance, for our specific tile decorations, of the cluster coverings proposed and analyzed for octagonal phases.~\cite{benabraham99prb,gaehler00mse,fu08ssc}

In advance of these future endeavors, we can draw some structural conclusions from our model.  It reaffirms the close structural relationship between octagonal quascrystals and the $\beta$-Mn structure inferred from experimental investigations.  In terms of local geometries, 8$_3$ helices play a prominent role in both, as Hovm{\"o}ller et al. assumed in their first structural model.~\cite{hovmoeller91pml}  However, a curious difference occurs in how these helices pack and interpenetrate to form the structures of octagonal phases and $\beta$-Mn. In $\beta$-Mn, symmetrically equivalent helices of this form propagate along $a$, $b$ and $c$, in accord with the structure's cubic symmetry.  These helices are tightly interpenetrating, with each atom lying simultaneously on at least two helices.

The octagonal phase structure is, however, decidedly uniaxial.  The structures of these materials, in our model, can be constructed by placing 8$_3$ helices at the vertices of a 2D octagonal tiling of squares and rhombi (in projection) with the same handedness and phase.  The remaining spaces are then filled to form unit cells of the $\beta$-Mn and rhombic phase structures for the square and rhombus tiles, respectively.  No reference is made to any sort of coupling between neighboring helices.  This uniaxial character is emphasized in our simulated electron diffraction patterns.  Our ideal structural model, designed to reproduce the results of diffraction experiments, shows an absence of reflection conditions connected with the 4$_1$ screw symmetry of these helices perpendicular to the octagonal axis.  An intriguing question is how this transformation from independent helices in quasicrystalline phases to tightly interconnecting helices in the $\beta$-Mn structure connects to the relative stabilities of these structures, and the kinetics for decomposition of octagonal phases into $\beta$-Mn-type ones.

Connections between octagonal phases and their twinned $\beta$-Mn-type decomposition products can also be seen on a larger length scale than their helical building units.  If, as we see in our simulations and was hypothesized by Kuo et al.,~\cite{Kuo87} the twinning in the experimentally observed $\beta$-Mn phases is mediated by small crystallites of the rhombic phase, then both the octagonal phase and their decomposition products consist of the same square and rhombic tiles.  Indeed, one could imagine this decomposition following an aggregation of square tiles into domains, with rhombic tiles segregating to the domain surfaces or merging to form new square tiles.   Viewing this decomposition in rewind mode, we can portray octagonal quasicrystals as derived from $\beta$-Mn via twinning on a progressively finer and finer length scale, until the domains consist of only one or two unit cells. We see then that in the case of octagonal phases there is a continuity between quasicrystals and twinned crystals; they represent the same interfacial phenomena occurring on different length scales.

\section*{Acknowledgments}

The authors gratefully acknowledge the support of the following Swedish Research Foundations: TFR, NFR and NTM and the Swedish National Infrastructure for Computing (SNIC) and the Centre for parallel Computers (PDC).  D.C.F. thanks the National Science Foundation for a post-doctoral fellowship (through Grant DMR-0502582).  Finally, we thank Dr. Junliang Sun for elucidating conversations regarding the simulation of electron diffraction patterns.

\appendix*
\section{Details of Potential Function and Molecular Dynamics Simulations}

Pair interaction appear to be sufficient for modelling liquid
metals.~\cite{Ashcroft90,Dzugutov94,Dzugutov89} The pair potential
used in the present study was constructed to imitate the interionic
interaction in simple metals, consisting of a short-range repulsive
core and a longer-range oscillatory part. The latter is meant to
represent the Friedel oscillations\cite{pettifor95book} which are
characteristic of the effective interionic potentials in simple
metals. The period of these oscillations is determined by the Fermi
wave-vector $K_F$ which is a function of the density of valence
electrons. The functional form of this potential is
\begin{equation}
  V(r)={A\exp(\alpha r)\cos(2K_{F}r) \over r^3} + Br^P+V_0
  \label{poteq}
\end{equation}
We used parameters for this potential which were previously determined
to encourage icosahedral arrangements in the first coordination
shell.~\cite{Doye03}$^,$~\footnote{Numbers used in simulation:
$A=1.58$, $B=0.95532948$, $K_F=4.12$, $\alpha=-0.22$, $P=-18$,
$V_0=0.04682632$.} This was achieved by putting an energy penalty at
the distance of $\sqrt{2}$ times that of the first potential minimum,
to discourage the formation of cubic structures. The same liquid, at a
higher density than that explored here, has also been observed to
crystallize in the $\gamma$-brass structure.\cite{Zetterling01}

\begin{figure}[h]
  \includegraphics[width=7.cm]{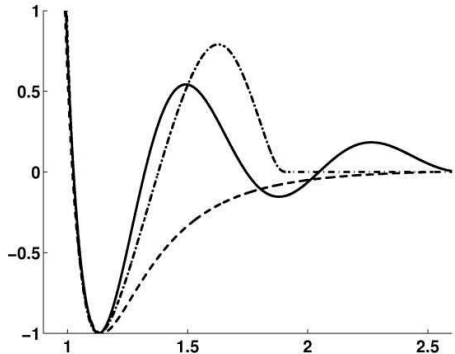}
  \caption{Pair-potentials, solid line:
    the current potential, dash-dotted line:
    IC potential \cite{Dzugutov92}, dotted line: Lennard-Jones
    potential. The energy level of the IC potential and the current potential
    has been scaled to give an energy of -1 at the first minimum.}
  \label{potplot}
\end{figure}

The potential is shown in Figure~\ref{potplot}, along with two other
well-known potentials for comparison, the IC potential of
Dzugutov~\cite{Dzugutov92} and the Lennard-Jones potential. The IC
potential also induces icosahedral short-range order and was used in
simulations of a dodecagonal quasicrystal~\cite{Dzugutov93} and a
$\sigma$-phase-type structure.~\cite{Sim00, Roth00} All three
potentials have nearly identical short-range repulsive parts. 

The simulations were performed using reduced units. The reduced units
correspond to those of the Lennard-Jones potential. In practice
this means that the length unit is defined as the onset of hard
repulsion, corresponding to half the particle radius. The mass unit is
the mass of a single particle. The energy unit is equal to the depth of
the first minimum of the potential and the time unit is derived from
the other three.

The simulation was performed at the constant number density
$\rho=0.84$ particles per unit volume with 16384 ($=16 \cdot 16 \cdot
16 \cdot 4$) particles. Newtonian equations were integrated using a
second order finite difference method, the leap-frog version of the
symplectic Verlet algorithm.\cite{Verlet67} Temperature adjustments
of the system were performed by scaling the velocities of the
particles. After each change in temperature an equilibration period
followed. Crystallization occurs at a temperature of T=0.45 in our
simulation units (note that the current potential in
Figure~\ref{potplot} is scaled. In that scale the crystallization
temperature is T=0.64). After crystallization, the resulting particle
configurations were treated with a steepest descent algorithm to
remove the static of thermal motions and yield the inherent
structure.\cite{Still84}


\begin{thebibliography}{41}
\expandafter\ifx\csname natexlab\endcsname\relax\def\natexlab#1{#1}\fi
\expandafter\ifx\csname bibnamefont\endcsname\relax
  \def\bibnamefont#1{#1}\fi
\expandafter\ifx\csname bibfnamefont\endcsname\relax
  \def\bibfnamefont#1{#1}\fi
\expandafter\ifx\csname citenamefont\endcsname\relax
  \def\citenamefont#1{#1}\fi
\expandafter\ifx\csname url\endcsname\relax
  \def\url#1{\texttt{#1}}\fi
\expandafter\ifx\csname urlprefix\endcsname\relax\def\urlprefix{URL }\fi
\providecommand{\bibinfo}[2]{#2}
\providecommand{\eprint}[2][]{\url{#2}}

\bibitem[{\citenamefont{Shechtman et~al.}(1984)\citenamefont{Shechtman, Blech,
  Gratias, and Cahn}}]{shechtman84prl}
\bibinfo{author}{\bibfnamefont{D.}~\bibnamefont{Shechtman}},
  \bibinfo{author}{\bibfnamefont{I.}~\bibnamefont{Blech}},
  \bibinfo{author}{\bibfnamefont{D.}~\bibnamefont{Gratias}}, \bibnamefont{and}
  \bibinfo{author}{\bibfnamefont{J.~W.} \bibnamefont{Cahn}},
  \bibinfo{journal}{Phys. Rev. Lett.} \textbf{\bibinfo{volume}{53}},
  \bibinfo{pages}{1951} (\bibinfo{year}{1984}).

\bibitem[{\citenamefont{Takakura et~al.}(2007)\citenamefont{Takakura,
  G{\'o}mez, Yamamoto, {de Boissieu}, and Tsai}}]{tsai07natmat}
\bibinfo{author}{\bibfnamefont{H.}~\bibnamefont{Takakura}},
  \bibinfo{author}{\bibfnamefont{C.~P.} \bibnamefont{G{\'o}mez}},
  \bibinfo{author}{\bibfnamefont{A.}~\bibnamefont{Yamamoto}},
  \bibinfo{author}{\bibfnamefont{M.}~\bibnamefont{{de Boissieu}}},
  \bibnamefont{and} \bibinfo{author}{\bibfnamefont{A.~P.} \bibnamefont{Tsai}},
  \bibinfo{journal}{Nat. Mater.} \textbf{\bibinfo{volume}{6}},
  \bibinfo{pages}{58} (\bibinfo{year}{2007}).

\bibitem[{\citenamefont{Katrych et~al.}(2007)\citenamefont{Katrych, Weber,
  Kobas, Mass{\"u}ger, Palatinus, Chapuis, and Steurer}}]{steurer07jac}
\bibinfo{author}{\bibfnamefont{S.}~\bibnamefont{Katrych}},
  \bibinfo{author}{\bibfnamefont{T.}~\bibnamefont{Weber}},
  \bibinfo{author}{\bibfnamefont{M.}~\bibnamefont{Kobas}},
  \bibinfo{author}{\bibfnamefont{L.}~\bibnamefont{Mass{\"u}ger}},
  \bibinfo{author}{\bibfnamefont{L.}~\bibnamefont{Palatinus}},
  \bibinfo{author}{\bibfnamefont{G.}~\bibnamefont{Chapuis}}, \bibnamefont{and}
  \bibinfo{author}{\bibfnamefont{W.}~\bibnamefont{Steurer}},
  \bibinfo{journal}{J. Alloys Compd.} \textbf{\bibinfo{volume}{428}},
  \bibinfo{pages}{164} (\bibinfo{year}{2007}).

\bibitem[{\citenamefont{Conrad et~al.}(1998)\citenamefont{Conrad, Krumeich, and
  Harbrecht}}]{harbrecht98ac}
\bibinfo{author}{\bibfnamefont{M.}~\bibnamefont{Conrad}},
  \bibinfo{author}{\bibfnamefont{F.}~\bibnamefont{Krumeich}}, \bibnamefont{and}
  \bibinfo{author}{\bibfnamefont{B.}~\bibnamefont{Harbrecht}},
  \bibinfo{journal}{Angew. Chem. Int. Ed.} \textbf{\bibinfo{volume}{37}},
  \bibinfo{pages}{1383} (\bibinfo{year}{1998}).

\bibitem[{\citenamefont{Conrad et~al.}(2000)\citenamefont{Conrad, Krumeich,
  Reich, and Harbrecht}}]{harbrecht00mse}
\bibinfo{author}{\bibfnamefont{M.}~\bibnamefont{Conrad}},
  \bibinfo{author}{\bibfnamefont{F.}~\bibnamefont{Krumeich}},
  \bibinfo{author}{\bibfnamefont{C.}~\bibnamefont{Reich}}, \bibnamefont{and}
  \bibinfo{author}{\bibfnamefont{B.}~\bibnamefont{Harbrecht}},
  \bibinfo{journal}{Mat. Sci. Eng.} \textbf{\bibinfo{volume}{294-296}},
  \bibinfo{pages}{37} (\bibinfo{year}{2000}).

\bibitem[{\citenamefont{Conrad and Harbrecht}(2002)}]{harbrecht02cej}
\bibinfo{author}{\bibfnamefont{M.}~\bibnamefont{Conrad}} \bibnamefont{and}
  \bibinfo{author}{\bibfnamefont{B.}~\bibnamefont{Harbrecht}},
  \bibinfo{journal}{Chem. Eur. J.} \textbf{\bibinfo{volume}{8}},
  \bibinfo{pages}{3095} (\bibinfo{year}{2002}).

\bibitem[{\citenamefont{Wang et~al.}(1987)\citenamefont{Wang, Chen, and
  Kuo}}]{Kuo87}
\bibinfo{author}{\bibfnamefont{N.}~\bibnamefont{Wang}},
  \bibinfo{author}{\bibfnamefont{H.}~\bibnamefont{Chen}}, \bibnamefont{and}
  \bibinfo{author}{\bibfnamefont{K.~H.} \bibnamefont{Kuo}},
  \bibinfo{journal}{Phys. Rev. Lett.} \textbf{\bibinfo{volume}{59}},
  \bibinfo{pages}{1010} (\bibinfo{year}{1987}).

\bibitem[{\citenamefont{Wang et~al.}(1988)\citenamefont{Wang, Fung, and
  Kuo}}]{kuo88apl}
\bibinfo{author}{\bibfnamefont{N.}~\bibnamefont{Wang}},
  \bibinfo{author}{\bibfnamefont{K.~K.} \bibnamefont{Fung}}, \bibnamefont{and}
  \bibinfo{author}{\bibfnamefont{K.~H.} \bibnamefont{Kuo}},
  \bibinfo{journal}{Appl. Phys. Lett.} \textbf{\bibinfo{volume}{52}},
  \bibinfo{pages}{2120} (\bibinfo{year}{1988}).

\bibitem[{\citenamefont{Steurer}(2004)}]{Steurer04}
\bibinfo{author}{\bibfnamefont{W.}~\bibnamefont{Steurer}}, \bibinfo{journal}{Z.
  Kristallographie} \textbf{\bibinfo{volume}{219}}, \bibinfo{pages}{391}
  (\bibinfo{year}{2004}).

\bibitem[{\citenamefont{Mai et~al.}(1989)\citenamefont{Mai, Xu, Wang, Kuo, Jin,
  and Cheng}}]{kuo89prb}
\bibinfo{author}{\bibfnamefont{Z.~H.} \bibnamefont{Mai}},
  \bibinfo{author}{\bibfnamefont{L.}~\bibnamefont{Xu}},
  \bibinfo{author}{\bibfnamefont{N.}~\bibnamefont{Wang}},
  \bibinfo{author}{\bibfnamefont{K.~H.} \bibnamefont{Kuo}},
  \bibinfo{author}{\bibfnamefont{Z.~C.} \bibnamefont{Jin}}, \bibnamefont{and}
  \bibinfo{author}{\bibfnamefont{G.}~\bibnamefont{Cheng}},
  \bibinfo{journal}{Phys. Rev. B} \textbf{\bibinfo{volume}{40}},
  \bibinfo{pages}{12183} (\bibinfo{year}{1989}).

\bibitem[{\citenamefont{Xu et~al.}(2000)\citenamefont{Xu, Wang, Lee, and
  Fung}}]{xu00prb}
\bibinfo{author}{\bibfnamefont{L.}~\bibnamefont{Xu}},
  \bibinfo{author}{\bibfnamefont{N.}~\bibnamefont{Wang}},
  \bibinfo{author}{\bibfnamefont{S.~T.} \bibnamefont{Lee}}, \bibnamefont{and}
  \bibinfo{author}{\bibfnamefont{K.~K.} \bibnamefont{Fung}},
  \bibinfo{journal}{Phys. Rev. B} \textbf{\bibinfo{volume}{62}},
  \bibinfo{pages}{3078} (\bibinfo{year}{2000}).

\bibitem[{\citenamefont{Socolar}(1989)}]{socolar89}
\bibinfo{author}{\bibfnamefont{J.~E.~S.} \bibnamefont{Socolar}},
  \bibinfo{journal}{Phys. Rev. B} \textbf{\bibinfo{volume}{39}},
  \bibinfo{pages}{10519} (\bibinfo{year}{1989}).

\bibitem[{\citenamefont{Ingalls}(1993)}]{ingalls93jncs}
\bibinfo{author}{\bibfnamefont{R.}~\bibnamefont{Ingalls}}, \bibinfo{journal}{J.
  Non-Cryst. Solids} \textbf{\bibinfo{volume}{153-154}}, \bibinfo{pages}{177}
  (\bibinfo{year}{1993}).

\bibitem[{\citenamefont{Wang and Kuo}(1988)}]{kuo88aca}
\bibinfo{author}{\bibfnamefont{Z.~M.} \bibnamefont{Wang}} \bibnamefont{and}
  \bibinfo{author}{\bibfnamefont{K.~H.} \bibnamefont{Kuo}},
  \bibinfo{journal}{Acta Cryst. A} \textbf{\bibinfo{volume}{44}},
  \bibinfo{pages}{857} (\bibinfo{year}{1988}).

\bibitem[{\citenamefont{Li and Cheng}(1996)}]{li96cpl}
\bibinfo{author}{\bibfnamefont{F.}~\bibnamefont{Li}} \bibnamefont{and}
  \bibinfo{author}{\bibfnamefont{Y.}~\bibnamefont{Cheng}},
  \bibinfo{journal}{Chin. Phys. Lett.} \textbf{\bibinfo{volume}{13}},
  \bibinfo{pages}{199} (\bibinfo{year}{1996}).

\bibitem[{\citenamefont{Zijlstra}(2004)}]{zijlstra04}
\bibinfo{author}{\bibfnamefont{E.~S.} \bibnamefont{Zijlstra}},
  \bibinfo{journal}{J. Non-Cryst. Solids} \textbf{\bibinfo{volume}{334-335}},
  \bibinfo{pages}{126} (\bibinfo{year}{2004}).

\bibitem[{\citenamefont{Liu et~al.}(1992)\citenamefont{Liu, Zhang, Jiang, and
  Tian}}]{liu92jpcm}
\bibinfo{author}{\bibfnamefont{Z.}~\bibnamefont{Liu}},
  \bibinfo{author}{\bibfnamefont{Z.}~\bibnamefont{Zhang}},
  \bibinfo{author}{\bibfnamefont{Q.}~\bibnamefont{Jiang}}, \bibnamefont{and}
  \bibinfo{author}{\bibfnamefont{D.}~\bibnamefont{Tian}}, \bibinfo{journal}{J.
  Phys.-Condens. Mat.} \textbf{\bibinfo{volume}{4}}, \bibinfo{pages}{6343}
  (\bibinfo{year}{1992}).

\bibitem[{\citenamefont{Kuo}(1990)}]{kuo90jncs}
\bibinfo{author}{\bibfnamefont{K.~H.} \bibnamefont{Kuo}}, \bibinfo{journal}{J.
  Non-Cryst. Solids} \textbf{\bibinfo{volume}{117/118}}, \bibinfo{pages}{756}
  (\bibinfo{year}{1990}).

\bibitem[{\citenamefont{Jiang et~al.}(1991)\citenamefont{Jiang, Wang, Fung, and
  Kuo}}]{kuo91prl}
\bibinfo{author}{\bibfnamefont{J.~C.} \bibnamefont{Jiang}},
  \bibinfo{author}{\bibfnamefont{N.}~\bibnamefont{Wang}},
  \bibinfo{author}{\bibfnamefont{K.~K.} \bibnamefont{Fung}}, \bibnamefont{and}
  \bibinfo{author}{\bibfnamefont{K.~H.} \bibnamefont{Kuo}},
  \bibinfo{journal}{Phys. Rev. Lett.} \textbf{\bibinfo{volume}{67}},
  \bibinfo{pages}{1302} (\bibinfo{year}{1991}).

\bibitem[{\citenamefont{Huang and Hovm{\"o}ller}(1991)}]{hovmoeller91pml}
\bibinfo{author}{\bibfnamefont{Z.}~\bibnamefont{Huang}} \bibnamefont{and}
  \bibinfo{author}{\bibfnamefont{S.}~\bibnamefont{Hovm{\"o}ller}},
  \bibinfo{journal}{Philos. Mag. Lett.} \textbf{\bibinfo{volume}{64}},
  \bibinfo{pages}{83} (\bibinfo{year}{1991}).

\bibitem[{\citenamefont{Jiang et~al.}(1995)\citenamefont{Jiang, Hovm{\"o}ller,
  and Zou}}]{hovmoeller95pml}
\bibinfo{author}{\bibfnamefont{J.-C.} \bibnamefont{Jiang}},
  \bibinfo{author}{\bibfnamefont{S.}~\bibnamefont{Hovm{\"o}ller}},
  \bibnamefont{and} \bibinfo{author}{\bibfnamefont{X.-D.} \bibnamefont{Zou}},
  \bibinfo{journal}{Philos. Mag. Lett.} \textbf{\bibinfo{volume}{71}},
  \bibinfo{pages}{123} (\bibinfo{year}{1995}).

\bibitem[{\citenamefont{Dzugutov}(1993)}]{Dzugutov93}
\bibinfo{author}{\bibfnamefont{M.}~\bibnamefont{Dzugutov}},
  \bibinfo{journal}{Phys. Rev. Lett.} \textbf{\bibinfo{volume}{70}},
  \bibinfo{pages}{2924} (\bibinfo{year}{1993}).

\bibitem[{\citenamefont{Quandt and Teter}(1999)}]{Quandt99}
\bibinfo{author}{\bibfnamefont{A.}~\bibnamefont{Quandt}} \bibnamefont{and}
  \bibinfo{author}{\bibfnamefont{M.~P.} \bibnamefont{Teter}},
  \bibinfo{journal}{Phys. Rev. B} \textbf{\bibinfo{volume}{59}},
  \bibinfo{pages}{8586} (\bibinfo{year}{1999}).

\bibitem[{\citenamefont{Engel and Trebin}(2007)}]{engel07prl}
\bibinfo{author}{\bibfnamefont{M.}~\bibnamefont{Engel}} \bibnamefont{and}
  \bibinfo{author}{\bibfnamefont{H.-R.} \bibnamefont{Trebin}},
  \bibinfo{journal}{Phys. Rev. Lett.} \textbf{\bibinfo{volume}{98}},
  \bibinfo{pages}{225505} (\bibinfo{year}{2007}).

\bibitem[{\citenamefont{Keys and Glotzer}(2007)}]{keys07prl}
\bibinfo{author}{\bibfnamefont{A.~S.} \bibnamefont{Keys}} \bibnamefont{and}
  \bibinfo{author}{\bibfnamefont{S.~C.} \bibnamefont{Glotzer}},
  \bibinfo{journal}{Phys. Rev. Lett.} \textbf{\bibinfo{volume}{99}},
  \bibinfo{pages}{235503} (\bibinfo{year}{2007}).

\bibitem[{\citenamefont{Hyde and Andersson}(1989)}]{anderssonbook}
\bibinfo{author}{\bibfnamefont{B.~G.} \bibnamefont{Hyde}} \bibnamefont{and}
  \bibinfo{author}{\bibfnamefont{S.}~\bibnamefont{Andersson}},
  \emph{\bibinfo{title}{Inorganic Crystal Structures}}
  (\bibinfo{publisher}{John Wiley and Sons}, \bibinfo{address}{New York},
  \bibinfo{year}{1989}).

\bibitem[{\citenamefont{Duneau et~al.}(1989)\citenamefont{Duneau, Mosseri, and
  Oguey}}]{Duneau89}
\bibinfo{author}{\bibfnamefont{M.}~\bibnamefont{Duneau}},
  \bibinfo{author}{\bibfnamefont{R.}~\bibnamefont{Mosseri}}, \bibnamefont{and}
  \bibinfo{author}{\bibfnamefont{C.}~\bibnamefont{Oguey}}, \bibinfo{journal}{J.
  Phys. A: Math. Gen.} \textbf{\bibinfo{volume}{22}}, \bibinfo{pages}{4549}
  (\bibinfo{year}{1989}).

\bibitem[{\citenamefont{Ben-Abraham and G{\"a}hler}(1999)}]{benabraham99prb}
\bibinfo{author}{\bibfnamefont{S.~I.} \bibnamefont{Ben-Abraham}}
  \bibnamefont{and}
  \bibinfo{author}{\bibfnamefont{F.}~\bibnamefont{G{\"a}hler}},
  \bibinfo{journal}{Phys. Rev. B} \textbf{\bibinfo{volume}{60}},
  \bibinfo{pages}{860} (\bibinfo{year}{1999}).

\bibitem[{\citenamefont{G{\"a}hler}(2000)}]{gaehler00mse}
\bibinfo{author}{\bibfnamefont{F.}~\bibnamefont{G{\"a}hler}},
  \bibinfo{journal}{Mater. Sci. Eng.} \textbf{\bibinfo{volume}{294-296}},
  \bibinfo{pages}{199} (\bibinfo{year}{2000}).

\bibitem[{\citenamefont{Liao and Fu}(2008)}]{fu08ssc}
\bibinfo{author}{\bibfnamefont{L.}~\bibnamefont{Liao}} \bibnamefont{and}
  \bibinfo{author}{\bibfnamefont{X.}~\bibnamefont{Fu}}, \bibinfo{journal}{Solid
  State Commun.} \textbf{\bibinfo{volume}{146}}, \bibinfo{pages}{35}
  (\bibinfo{year}{2008}).

\bibitem[{\citenamefont{Dzugutov}(1990)}]{Ashcroft90}
\bibinfo{author}{\bibfnamefont{N.~W.}~\bibnamefont{Ashcroft}},
  \bibinfo{journal}{Nuovo Cimento Soc. Ital. Fis., D} 
  \textbf{\bibinfo{volume}{12}},
  \bibinfo{pages}{597} (\bibinfo{year}{1990}).

\bibitem[{\citenamefont{Dzugutov}(1989)}]{Dzugutov89}
\bibinfo{author}{\bibfnamefont{M.}~\bibnamefont{Dzugutov}},
  \bibinfo{journal}{Phys. Rev. A} \textbf{\bibinfo{volume}{40}},
  \bibinfo{pages}{5434} (\bibinfo{year}{1989}).

\bibitem[{\citenamefont{Dzugutov, Alvarez, and Lomba}(1994)}]{Dzugutov94}
\bibinfo{author}{\bibfnamefont{M.}~\bibnamefont{Dzugutov}},
  \bibinfo{author}{\bibfnamefont{M.}~\bibnamefont{Alvarez}},
  \bibnamefont{and} \bibinfo{author}{\bibfnamefont{E.}~\bibnamefont{Lomba}},
  \bibinfo{journal}{J. Phys.: Condens. Matter} \textbf{\bibinfo{volume}{6}},
  \bibinfo{pages}{4419} (\bibinfo{year}{1994}).


\bibitem[{\citenamefont{Pettifor}(1995)}]{pettifor95book}
\bibinfo{author}{\bibfnamefont{D.~G.} \bibnamefont{Pettifor}},
  \emph{\bibinfo{title}{Bonding and Structure of Molecules and Solids}}
  (\bibinfo{publisher}{Oxford University Press}, \bibinfo{address}{Oxford},
  \bibinfo{year}{1995}).

\bibitem[{\citenamefont{Doye et~al.}(2003)\citenamefont{Doye, Wales,
  Zetterling, and Dzugutov}}]{Doye03}
\bibinfo{author}{\bibfnamefont{J.~P.~K.} \bibnamefont{Doye}},
  \bibinfo{author}{\bibfnamefont{D.~J.} \bibnamefont{Wales}},
  \bibinfo{author}{\bibfnamefont{F.~H.~M.} \bibnamefont{Zetterling}},
  \bibnamefont{and} \bibinfo{author}{\bibfnamefont{M.}~\bibnamefont{Dzugutov}},
  \bibinfo{journal}{J. Chem. Phys.} \textbf{\bibinfo{volume}{118}},
  \bibinfo{pages}{2792} (\bibinfo{year}{2003}).

\bibitem[{\citenamefont{Zetterling et~al.}(2001)\citenamefont{Zetterling,
  Dzugutov, and Lidin}}]{Zetterling01}
\bibinfo{author}{\bibfnamefont{F.}~\bibnamefont{Zetterling}},
  \bibinfo{author}{\bibfnamefont{M.}~\bibnamefont{Dzugutov}}, \bibnamefont{and}
  \bibinfo{author}{\bibfnamefont{S.}~\bibnamefont{Lidin}},
  \bibinfo{journal}{MRS Symposium Proc.} \textbf{\bibinfo{volume}{643}},
  \bibinfo{pages}{K9.5.1} (\bibinfo{year}{2001}).

\bibitem[{\citenamefont{Dzugutov}(1992)}]{Dzugutov92}
\bibinfo{author}{\bibfnamefont{M.}~\bibnamefont{Dzugutov}},
  \bibinfo{journal}{Phys. Rev. A} \textbf{\bibinfo{volume}{46}},
  \bibinfo{pages}{R2984} (\bibinfo{year}{1992}).

\bibitem[{\citenamefont{Simdyankin et~al.}(2000)\citenamefont{Simdyankin,
  Taraskin, Dzugutov, and Elliott}}]{Sim00}
\bibinfo{author}{\bibfnamefont{S.~I.} \bibnamefont{Simdyankin}},
  \bibinfo{author}{\bibfnamefont{S.~N.} \bibnamefont{Taraskin}},
  \bibinfo{author}{\bibfnamefont{M.}~\bibnamefont{Dzugutov}}, \bibnamefont{and}
  \bibinfo{author}{\bibfnamefont{S.~R.} \bibnamefont{Elliott}},
  \bibinfo{journal}{Phys. Rev. B} \textbf{\bibinfo{volume}{62}},
  \bibinfo{pages}{3223} (\bibinfo{year}{2000}).

\bibitem[{\citenamefont{Roth and Denton}(2000)}]{Roth00}
\bibinfo{author}{\bibfnamefont{J.}~\bibnamefont{Roth}} \bibnamefont{and}
  \bibinfo{author}{\bibfnamefont{A.~R.} \bibnamefont{Denton}},
  \bibinfo{journal}{Phys. Rev. E} \textbf{\bibinfo{volume}{61}},
  \bibinfo{pages}{6845} (\bibinfo{year}{2000}).

\bibitem[{\citenamefont{Verlet}(1967)}]{Verlet67}
\bibinfo{author}{\bibfnamefont{L.}~\bibnamefont{Verlet}},
  \bibinfo{journal}{Phys. Rev.} \textbf{\bibinfo{volume}{159}},
  \bibinfo{pages}{98} (\bibinfo{year}{1967}).

\bibitem[{\citenamefont{Stillinger and Weber}(1984)}]{Still84}
\bibinfo{author}{\bibfnamefont{F.~H.} \bibnamefont{Stillinger}}
  \bibnamefont{and} \bibinfo{author}{\bibfnamefont{T.~A.} \bibnamefont{Weber}},
  \bibinfo{journal}{Science} \textbf{\bibinfo{volume}{225}},
  \bibinfo{pages}{983} (\bibinfo{year}{1984}).

\end{thebibliography}
\end{document}